\documentclass[11pt,aps,prd,amsfonts,nofootinbib,supercriptaddress]{revtex4}
\usepackage{amssymb,amsmath}
\usepackage{graphicx}
\usepackage{epsfig}

\newcommand{\beq}{\begin{equation}}
\newcommand{\eeq}{\end{equation}}
\newcommand{\bea}{\begin{eqnarray}}
\newcommand{\eea}{\end{eqnarray}}

\begin{document}

\title{Radial stability analysis of the continuous pressure gravastar}

\author{Dubravko Horvat~\footnote{dubravko.horvat@fer.hr}, Sa\v sa Iliji\' c~\footnote{sasa.ilijic@fer.hr},
Anja Marunovi\' c~\footnote{anja.marunovic@fer.hr}}
\affiliation{Department of Physics, Faculty of Electrical
Engineering and Computing, University of Zagreb, Unska 3, HR-10
000 Zagreb, Croatia}

\begin{abstract} \noindent

 Radial stability of the continuous pressure gravastar is studied using the conventional
 Chandrasekhar method. The equation of state for the static gravastar solutions is derived
 and Einstein equations for small perturbations around the equilibrium are solved
 as an eigenvalue problem for radial pulsations.
Within the model there exist a set of parameters leading
 to a stable fundamental mode, thus proving radial stability of the continuous pressure
 gravastar. It is also shown that the central energy
  density possesses an
 extremum in $\rho_c(R)$ curve which represents a splitting point between stable and unstable
 gravastar configurations. As such the $\rho_c(R)$ curve for the gravastar mimics the famous $M(R)$ curve for
 a polytrope. Together with the former axial stability calculations this work completes the
 stability problem of the continuous pressure gravastar.

\end{abstract}

\maketitle

\section{Introduction}

Gravitational collapse as the stellar nuclear fuel is consumed
could lead to black holes - objects which are accepted by
scientific community  but their undesired  and even paradoxical
features (singularities, horizon) have motivated research in a
direction of finding massive objects (stars) without singularities
and without horizon.  One of these alternatives is a gravastar.

Since the seminal work of Mazur and Mottola \cite{MazurMottola} the concept
of the \textbf{gra}vitational \textbf{va}cuum \textbf{star}
-- the gravastar -- as an alternative to a
black hole has attracted a plethora of interest.
In this  version of the gravastar a  multilayered structure has
been introduced:
from the repulsive de Sitter core (where a negative pressure helps
balance the collapsing matter) one crosses multiple layers (shells)
and without encountering a horizon one eventually reaches the
(pressureless) exterior Schwarzschild spacetime.
Later some simplifications \cite{Carter,Visser} and modifications
\cite{Lobo:StableDES} have been
introduced in the original (multi)layer - onion-like picture.

An important step was done when it was shown that due to
anisotropy of matter comprising the gravastar \cite{Cattoen} one
can eliminate layer(s) and, by a continuous stress-energy tensor,
the transition from the interior de Sitter spacetime segment to
the exterior Schwarzschild spacetime is possible
\cite{DeBenedictis} (see also \cite{Dymnikova}).
The pressure anisotropy in the spherically symmetric geometry was perhaps first
introduced by G. Lema\^\i tre \cite{Lemaitre} and suggested by
Einstein (as quoted in \cite{Lemaitre}). The vanishing radial
pressure with transversal
pressure only was shown to be enough to
support a stable object. Further development \cite{Bowers-Liang,
Fuzfa-etal} has brought different
refinements to the original anisotropy notion.
The pressure anisotropy, which is shown to be a necessary
condition for the existence of a gravastar \cite{Cattoen} is met
also in boson star models \cite{Schunck-Mielke}
and wormholes \cite{Garattini}. The anisotropy (defined as a difference
between the transversal and radial pressure)
vanishes  at the center ($r=0$) of the star as well as at its
boundary ($r=R$).
The gravastar has
been confronted with  its rivals - black holes
\cite{DeBenedictisRev,RochaMCh} and wormholes
\cite{Garattini,SushkovZaslavski,DeBenedictisGarLobo}, and
investigated with respect to energy conditions (violations)
\cite{HorvatEC} and its charged properties
\cite{Horvat:Charged,LoboArellano}. An interesting question has
been posed several times: is it possible to distinguish the
gravastar from a black hole
\cite{HarkoKL,Rezzolla,BroderickNarayan}? In Ref. \cite{Rezzolla} it was shown that
gravitational radiation could be used to tell a gravastar from a black hole.
However, the definite answer to this question has
not been given at satisfactory level and the gravastar research is still a dynamic field
with recent papers like
\cite{Lobo:Noncommutative,PaniBerti,BrandtChan,UsmaniRahaman}
\emph{etc}.

Almost every research mentioned above to some extent
 addresses the problem of the gravastar stability, since the stability problem is
crucial for any object or situation to be considered as physically
viable. In Ref. \cite{MazurMottola} it was first shown that such
an object is thermodynamically stable,  while axial stability of
thin-shells gravastars was tested in \cite{Carter,Lobo:StableDES}.
Stability within the thin shell approach based on the
Darmois-Israel formalism was recently reviewed in
\cite{LoboCrawford}. In \cite{Rezzolla} stability analysis of the
thin shell gravastar problem is closely related to an attempt to
distinguish the gravastar from a black hole by analysis of
quasi-normal modes produced by axial perturbations. Problem of
stability of a rotating the thin shell gravastar was addressed in
\cite{ChirentiRezzolla}. Stability in the (multi)layer version of
the gravastar was also considered in
\cite{GasparRacz,PaniBerti,PaniCardoso,RochaMCh}.

The axial stability of the continuous pressure gravastar was shown
to be valid in \cite{DeBenedictis}. This analysis was based on the
Ref. \cite{DymnikovaGalaktionov} where stability of objects with
de Sitter centers was investigated.

 In this paper we analyze the
 radial stability of the continuous pressure gravastars
 \cite{DeBenedictis}  following the
 conventional Chandrasekhar method.
 Originally Chandrasekhar developed the method
 for testing the radial stability of the isotropic spheres
 \cite{Chandrasekhar:1964} in terms of the radial
 pulsations. In Ref. \cite{Gleiser:Stability}
 Chandrasekhar's method was generalized to  anisotropic spheres.
Stability of anisotropic stars was investigated before in
\cite{HeintzmannHillebrandt,Hillebrandt}   and
radial stability analysis for anisotropic stars
using the quasi-local equation of state  was given in \cite{Horvat:Stability}.
The standard mathematical procedure is here applied to an object with
a peculiar behavior
of pressures (see below) and although the mathematical rigor was never
abandoned, the
analysis due to the character  of the object could be considered as a
toy model analysis of radial stability.

This paper is organized as follows. In the next section the
linearization of the Einstein equations is given. Static solutions
are described, an equation of state is derived and the pulsation
equation is obtained. In Sec. III the eigenvalue problem for the
radial displacement is presented. Results and discussion are given
in the
last section.\\
Unless stated explicitly we shall work in units where
 $G_N=1=c$.

\section{Linearization of the Einstein equation}

In this paper the response of the continuous pressure gravastar model to small
radial perturbations is considered. Assuming that the pulsating
object retains its spherical symmetry, one can introduce the
Schwarzschild coordinates:
\beq
ds^2=e^{\nu(r,t)}dt^2-e^{\lambda(r,t)}dr^2-r^2d\theta^2-r^2\sin^2\theta d\phi^2,
\label{metric}
\eeq
where $\lambda$ and $\nu$ are, in this dynamical setting, time-dependent metric functions.\\
The standard anisotropic energy-momentum tensor appropriate to describe continuous pressure gravastars is:
\beq
T_\mu^\nu=(\rho +p_r)u_\mu u^\nu-g_\mu^\nu p_r+l_\mu l^\nu(p_t-p_r)
   +k_\mu k^\nu(p_t-p_r),
\label{e-m tensor}
\eeq
where $u^\mu$ is the fluid 4-velocity, $u^\mu=dx^\mu /ds$, $l_\mu$ and $k_\mu$ are the unit
4-vectors in the $\theta$ and $\phi$ directions, respectively, $l_\mu =-r \,\delta_\mu^\theta,\;
 k_\mu = -r\sin\theta\,\delta_\mu^\phi$.\\
The velocity of the fluid element in the radial direction
$\dot\xi$ is defined by:
\beq \dot{\xi}\equiv \frac{dr}{dt}=\frac{u^r}{u^t},
\label{velocity} \eeq
where $\xi$ is the radial displacement of the fluid element,
$r\rightarrow r+\xi(r,t)$.
The components of the 4-velocity are obtained by employing  $u_\mu
u^\mu=1$ and Eq.~(\ref{velocity}):
\beq u^\mu=(e^{-\nu/2},\dot \xi
e^{-\nu/2},0,0). \label{components-velocity} \eeq
The non-zero components of the energy-momentum tensor~(\ref{e-m
tensor}) linear in $\dot{\xi}$ are:
\beq T_t^t = \rho,\quad T_r^r = -p_r, \quad T_\theta^\theta =
T_\phi^\phi=-p_t,\quad T_t^r = \dot \xi (\rho+p_r), \quad
T_r^t=e^{\lambda-\nu}\dot\xi (\rho+p_r). \label{en-mom tensor
components} \eeq
The components of the Einstein tensor for the
metric~(\ref{metric}) are:
\bea G_t^t & = &
e^{-\lambda}\left(\frac{\lambda^\prime}{r}-\frac{1}{r^2}\right)+\frac{1}{r^2},
\label{Gtt}\\
G_r^r & = &
-e^{-\lambda}\left(\frac{\nu^\prime}{r}+\frac{1}{r^2}\right)+
\frac{1}{r^2}, \label{Grr}\\
G_t^r & = &-e^{-\lambda}\frac{\dot{\lambda}}{r},
 \label{Gtr}\\
G_\theta^\theta=G_\phi^\phi &=&-\frac{1}{2}e^{-\lambda} \left(
-\frac{\nu^\prime \lambda^\prime}{2}
-\frac{\lambda^\prime}{r}+\frac{\nu^\prime}{r}+\frac{\nu^{\prime
2}}{2}+\nu^{\prime \prime} \right) + \frac{1}{2}e^{-\nu} \left(
\ddot{\lambda}+\frac{\dot{\lambda
^2}}{2}-\frac{\dot{\lambda}\dot{\nu}}{2}  \right). \label{G-theta}
\eea
Following the standard Chandrasekhar method, all matter and
metric functions should only slightly deviate
 from their equilibrium solutions,
\beq
\lambda(r,t) = \lambda_0(r)+\delta \lambda (r,t),\;\nu(r,t)=\nu_0(r)+\delta \nu (r,t),
\eeq
\beq
\rho(r,t)=\rho_0(r)+\delta \rho (r,t),\;
p_r(r,t)=p_{r0}(r)+\delta p_r (r,t),
p_t(r,t)=p_{t0}(r)+\delta p_t (r,t).
\eeq
The subscript $0$ denotes the equilibrium functions and $\delta
f(r,t)$ are the so-called Eulerian perturbations, where
$f\in\{\lambda,\nu,\rho,p_r,p_t,\}$. The Eulerian perturbations
measure a local departure from equilibrium in contrast to the
Lagrangian perturbations, denoted as $df(r,t)$, which measure a
departure from equilibrium in the co-moving system (fluid rest
frame). The Lagrangian perturbations in the linear approximation
play a role of a total differential and are linked to the Eulerian
perturbations via the equation (see \emph{e.g.} Ref.~\cite{MTW}):
\beq df(r,t)=\delta f(r,t)+f_0^\prime (r)\xi.
 \label{Lagr-Euler}
\eeq
A linearization of the Einstein equations $G_\mu^\nu=8\pi
T_\mu^\nu$ leads to the two sets of equations: one for the
equilibrium (static) functions and the other for the perturbed
functions. The equilibrium functions obey the following set of
equations:
\bea
8\pi\rho_0&=&e^{-\lambda_0}\left(\frac{\lambda_0^\prime}{r}-\frac{1}{r^2}\right)+\frac{1}{r^2},\\
\label{Gtt TOV}
8\pi p_{r0}&=&e^{-\lambda_0}\left(\frac{\nu_0^\prime}{r}+\frac{1}{r^2}\right)-\frac{1}{r^2},\\
\label{Grr TOV} 8\pi p_{t0}&=&\frac{1}{2}e^{-\lambda_0} \left(
-\frac{\nu_0^\prime \lambda_0^\prime}{2}
-\frac{\lambda_0^\prime}{r}+\frac{\nu_0^\prime}{r}+\frac{\nu_0^{\prime
2}}{2}+\nu_0^{\prime \prime} \right). \label{G theta TOV} \eea
In practice, one usually  combines these three equations into the
Tolman-Oppenheimer-Volkoff (TOV) equation:
\beq p_{r0}^\prime=-\frac 12
\left(\rho_0+p_{r0}\right)\nu_0^\prime +\frac{2}{r}\Pi_0,
\label{TOV} \eeq
where $\Pi_0$ denotes the anisotropic term $\Pi_0=p_{t0}-p_{r0}$.
The other set of equations emerging from the linearization of the
above Einstein equations yield the set of equations for the
perturbed functions:
\beq \left( re^{-\lambda_0}\delta \lambda \right)^\prime=8\pi r^2
\delta \rho, \label{first eq} \eeq
\beq \delta \nu ^\prime=\left( \nu_0^\prime+\frac{1}{r}\right)
\delta \lambda+8\pi r e^{\lambda_0}\delta p_r, \label{second eq}
\eeq
\beq \dot{\delta\lambda}
\frac{e^{-\lambda_0}}{r}=-8\pi\dot\xi(\rho_0+p_{r0}), \label{third
eq} \eeq
\beq
e^{\lambda_0-\nu_0}(\rho_0+p_{r0})\ddot\xi+\frac{1}{2}(\rho_0+p_{r0})\delta\nu^\prime +\frac{1}{2}(\delta\rho+\delta p_{r})\nu_0^\prime+\delta p_r^\prime-\frac{2}{r}\delta \Pi=0.
\label{pulsation eq}
\eeq
Equation~(\ref{pulsation eq}) is known as the \emph{pulsation
equation} \cite{Gleiser:Stability} and it serves to probe the
radial stability of the system of interest. It is actually the TOV
equation for the perturbed functions which is obtained --
analogously as the non-perturbed TOV -- by combining Einstein
equations for perturbed functions. \\
In order to solve the pulsation equation~(\ref{pulsation eq}) for
the gravastar all perturbed functions should be expressed
in terms of the radial displacement $\xi$ (and its derivatives)
and the equilibrium functions. In performing this, one first
integrates Eq.~(\ref{third eq}) in time, yielding:
\beq
\delta \lambda=-8\pi r e^{\lambda_0}\xi(\rho_0+p_{r0}).
\label{d lambda}
\eeq
Using this expression for $\delta \rho$ in Eq.~(\ref{first eq})
one obtains:
\beq
\delta\rho=-\frac{1}{r^2}\left[r^2(\rho_0+p_{r0})\xi\right]^\prime.
\label{d rho}
\eeq
After inserting $\delta \lambda$ in Eq.~(\ref{second eq}) a
dependence on $\delta p_r$ remains, which should be expressed in
terms of the displacement function (and its derivatives) and the
equilibrium functions. To accomplish this, one ought to explore
 the system at hand in more detail.

One of the possibilities, as suggested firstly by Chandrasekhar
for isotropic structures~\cite{Chandrasekhar:1964} and more
recently by Dev and Gleiser for anisotropic
objects~\cite{Gleiser:Stability}, is to make use of the baryon
density conservation to express the radial pressure perturbation
in terms of the displacement function and the static solutions. In
this approach the adiabatic index appears as a free parameter.
Chandrasekhar used this method to establish limiting values of the
adiabatic index leading to an (un)stable isotropic object of a
constant energy density. He showed that there were no stable stars
of this kind if the adiabatic index was less than $4/3+\kappa M/R$
($\kappa$ is a constant of order unity depending on the structure
of the star, $M$ and $R$ are the star's mass and radius). In
Ref.~\cite{Gleiser:Stability} the Chandrasekhar method was
extended to various anisotropic star models and showed that the
limiting value of the adiabatic index is shifted to lower values,
\emph{i.e.} anisotropic stars can approach the
 stability region
 with smaller adiabatic index than in the Chandrasekhar's case.

 In this paper our primary concern is to probe the radial stability
 of one particular anisotropic object -- the gravastar.
  Due to the peculiar character of the gravastar (especially its
  radial pressure -- see below)  one cannot expect the
adiabatic index to be constant along the whole object. In fact the
adiabatic index is a function of the energy density and
pressure(s). This is the main reason why in this paper stability will not
be tested by fixing the appropriate values of the
adiabatic index that guarantee stability. The required information
will rather be extracted from a given static solution by
constructing the equation of state.

\subsection{Static solution}
The procedure discussed so far is applicable to all spherically
symmetric structures. To apply it to gravastar configurations one
has to recall the basic characteristics of gravastars in the
continuous pressure picture~\cite{DeBenedictis}. The energy
density $\rho_0(r)$ is positive and monotonically decreases from
the center to the surface. Gravastars have a de Sitter-like
interior, $p_{r0}(0)=-\rho_0(0)$, and a Schwarzschild-like
exterior. Furthermore, the atmosphere of the gravastar is defined
as an outer region, near to the surface, where "normal" physics is
valid~\cite{Cattoen}, \emph{i.e.} where both the energy-density
and the radial pressure are positive and monotonically decreasing
functions of the radius. In the gravastar's atmosphere the sound
velocity $v_s$, with
\beq v_s^2=\frac{dp_{r0}}{d\rho_0},
\label{sound velocity} \eeq
is real ($v_s^2>0$) and subluminal ($v_s<1$).\\
From the peculiar shape of the gravastar's (radial) pressure
one can immediately infer that the sound velocity ought to be real
only in the gravastar's atmosphere, whilst in the gravastar's
interior it is imaginary, $v_s^2 < 0$. This is the main reason
why, in probing the radial stability, we shall be primarily concerned
with the physical processes occurring in the gravastar's
atmosphere.

To construct a static gravastar, the energy density profile and
the anisotropic term are adopted from the previous
work~\cite{DeBenedictis,Horvat:Charged}:
\beq \rho_0 (r)=\rho_c(1-(r/R)^n),
\label{en-density} \eeq
\beq
 \Pi_0
(r)=\beta \rho_0(r)^m \mu_0 (r). \label{anisotropy}
 \eeq
Here $n$, $m$ are (free) parameters and $\rho_c=\rho_0(0)$ is the
central energy density. $\beta$ is the anisotropy strength measure
and $R$ is the radius of the gravastar for which $p_{r0}(R)=0$.
$\mu_0(r)$ is the compactness function defined by $\mu_0(r)=2
m_0(r)/r$, where $m_0(r)$ is the mass function $m_0(r)=4\pi\int
\rho_0(r) r^2 dr$. The radial pressure $p_{r0}$ is a solution of
the TOV~(\ref{TOV}) and the tangential pressure is readily
obtained from the anisotropy and the radial pressure by employing
the identity $p_{t0}=p_{r0}+\Pi_0$. The particular form of
the anisotropic term is dictated by the
behavior of pressures at
de Sitter core, since at $r=0$ the anisotropy should vanish as seen
from (16). Also, the above
form of
the anisotropy term ensures that the radial pressure vanishes at
$r=R$. The transversal
pressure vanishes as well although it is not necessary to be the case
(see Ref. \cite{DeBenedictis}
for the  gravastar model with  non vanishing transverse  pressure.). The
anisotropy strength measure
is controlled as well by the energy conditions which have to be met.

\begin{figure}
\begin{center}
\leavevmode
\includegraphics[scale=0.9]{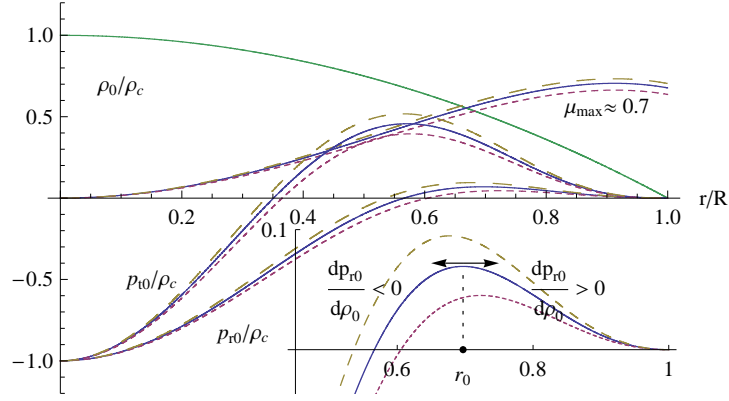}
\end{center}
\caption{The energy density $\rho_0/\rho_c$, radial pressure
$p_{r0}/\rho_c$, tangential pressure $p_{t0}/\rho_c$  and
compactness $\mu_0$ as a function of radius $r/R$ for
$\{R,n,m\}=\{1,2,3\}$. Three different values of the central
energy density $\rho_c=\{0.190,0.202,0.210\}$ and their anisotropy
strengths $\beta=\{92.905,81.410,76.110\}$ correspond to the
lower, middle and upper curve, respectively. $r_0$ denotes the
radius at which the sound velocity~(\ref{sound velocity}) vanishes
(for the central curve).}
 \label{fig1}
\end{figure}

One such solution for fixed $(R,n,m)=(1,2,3)$ is shown in Fig. 1
for three different values of the central energy density $\rho_c$
corresponding to three different values of the anisotropy strength
 $\beta$. Since the radius $R$ is fixed there is an
interplay between the central energy density $\rho_c$ and
anisotropy strength $\beta$ -- a higher central energy density
$\rho_c$ requires a smaller anisotropy strength $\beta$. We shall
elaborate on this particular choice of parameters in Section IV,
where the radial stability of these three gravastar configurations
will be tested.\\
In the inset of Fig. 1 the radial pressure close to the surface is
extracted in order to show important features of the gravastar's
atmosphere. At the radius $r_0$ the sound velocity of the fluid
vanishes ($dp_{r0}/d\rho_0|_{r=r_0}=0$) and hence $r_0$ serves as
a division point of the propagating (or physically reasonable)
($r>r_0$, $v_s^2>0$) and non-propagating regions ($r<r_0$,
$v_s^2<0$) when
probing radial pulsations of the gravastar.\\
The dominant energy condition (DEC), \emph{i.e.} $p_{r0},
p_{t0}\le\rho_0$, is obeyed by both radial and tangential pressure
throughout the gravastar. The compactness function has also been
shown in Fig. 1.
\subsection{Equation of state}

In this subsection we note that the equation of state (EoS)
appropriate to describe the gravastar (inferred from the input
functions~(\ref{en-density}) and~(\ref{anisotropy})) is actually a
functional of the energy density (only), parameterized by the
anisotropy strength $\beta$. Next this result is used to compute
the Eulerian perturbation of the radial pressure $\delta p_r$ from
the EoS, by perturbing the energy density only. Ultimately this
completes the task to express all perturbed functions
in terms of the displacement (and its derivatives) and the static solutions. \\
Generally, for isotropic structures, before solving the TOV, one
assumes that the pressure $p$ and the energy density $\rho$ are
functions of the specific entropy $s$ and the baryon density $n$.
If a system is described by the one-fluid model, then in static
and dynamic settings it exhibits isentropic behavior (constant
$s$), in which case one can set $s=0$. Thus it is possible to
eliminate the baryon density $n$ and express the pressure in terms
of the energy density only, leading to a barotropic equation of
state $p=p(\rho)$.
It is a rather simple task now to perturb this EoS and express the perturbed
pressure in terms of the perturbed energy density.\\

\begin{figure}
\begin{center}
\leavevmode
\includegraphics[scale=0.9]{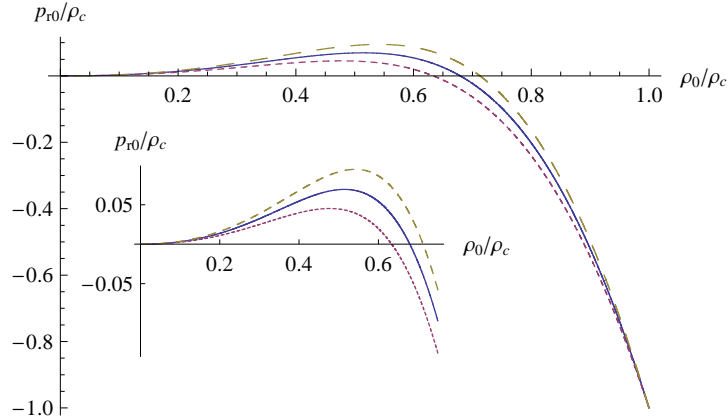}
\end{center}
\caption{The radial pressure $p_{r0}/\rho_c$ is plotted against
the energy density $\rho_0/\rho_c$ (EoS) for
$\{R,n,m\}=\{1,2,3\}$. Three different values of the central
energy density $\rho_c=\{0.190,0.202,0.210\}$ and their anisotropy
strengths $\beta=\{92.905,81.410,76.110\}$ correspond to the
lower, middle and upper curve, respectively.} \label{fig2}
\end{figure}

For anisotropic objects the EoS is highly dependent on the
anisotropic term model (see \emph{e.g.} the TOV~(\ref{TOV})). The
particular choice of the anisotropic term used
here~(\ref{anisotropy}) is a functional (or a quasi-local
variable~\footnote{By the quasi-local variable we mean a function
which is an integral in space of some local function -- for
example, the mass function $m_0(r)$ is a quasi-local variable of
the energy density (which is a local function) as it is the volume
integral of the energy density (the same holds for the compactness
function). For a discussion of quasi-local variables and
quasi-local EoS see \emph{e.g.}
Refs.~\cite{Hernandes:Eos,Hernandez:EoS-aniz} and
Ref.~\cite{Horvat:Stability}.}) of the energy density.
This means that for a fixed anisotropy strength $\beta$ there is a
two parameter family of values $\{\rho_c,R\}$ belonging to the
same EoS (see Fig. 2). As a consequence, one can obtain perturbed
(radial) pressure by perturbing the energy density only, and
keeping the anisotropy strength $\beta$ fixed.

To illustrate this in more detail let us introduce an analytic
form of the EoS which, to a good approximation, describes the
gravastar configuration defined by~(\ref{en-density})
and~(\ref{anisotropy}): ~\footnote{It is worth noting that the
analytic form of the EoS~(\ref{analytic EoS}) is not restricted to
the chosen energy density~(\ref{en-density}). For example, it is
also appropriate to describe a gravastar with the energy density
of the form $\rho_0(r)=\rho_c\, e^{-\eta\, r^2}$.}
\beq p_{r0}[\rho_0]=-\rho_0^2\left(\frac {1}{\rho_c}-\alpha\;
\mu_0[\rho_0]\right).
\label{analytic EoS}
\eeq
 Here $\alpha$ is closely related to the anisotropy strength $\beta$,
 $\mu_0[\rho_0]$ is the compactness function which is a functional
 of the energy density. Now it is clear that for a fixed $\alpha$
  the (radial) pressure is fully determined by the energy density.

 Hence, following the reasoning outlined above and making use of Eq.~(\ref{Lagr-Euler}), in the linear approximation
 the Eulerian perturbation for the radial pressure is:
\beq \delta
p_r=-p_{r0}^\prime\xi+\frac{dp_{r0}[\rho_0]}{d\rho_0}(\delta\rho+\rho_0^\prime\xi).
\label{delta p_r} \eeq
Here $dp_{ro}[\rho_0]/d\rho_0$ denotes functional derivative of
the radial pressure with respect to the energy density. This is
equal to $\frac{dp_{ro}/dr}{d\rho_0/dr}$ as both the radial
pressure and the
energy density are functions of radius $r$ only.\\
Similarly, the Eulerian perturbation of the anisotropy $\delta\Pi$
assumes the form:
\beq
\delta\Pi=-\Pi_0^\prime\xi+\frac{d\Pi_0[\rho_0]}{d\rho_0}(\delta\rho+\rho_0^\prime\xi).
\label{delta Pi}
 \eeq
With the above two expressions the pulsation
equation~(\ref{pulsation eq}) is fully determined.
However, before proceeding to solve the pulsation equation it is
useful to rewrite Eq.~(\ref{delta p_r}) in a slightly different
form in order to compare the result in this paper with that of
Chandrasekhar's for isotropic, and Dev and Gleiser's for
anisotropic stars. By means of the TOV~(\ref{TOV}) the perturbed
energy density~(\ref{d rho}) can be written as
\beq \delta\rho=-\rho_0^\prime
\xi-(\rho_0+p_{r0})\frac{e^{\nu_0/2}}{r^2}\left(r^2
e^{-\nu_0/2}\xi\right)^\prime-\frac 2 r \Pi_0\xi. \label{d rho 2}
\eeq
Inserting this result in Eq.~(\ref{delta p_r}) the radial pressure
perturbation becomes
\beq \delta
p_r=-p_{r0}^\prime\xi-(\rho_0+p_{r0})\frac{dp_{r0}[\rho_0]}{d\rho_0}\frac{e^{\nu_0/2}}{r^2}\left(r^2
e^{-\nu_0/2}\xi\right)^\prime-\frac 2 r   \Pi_0
\frac{dp_{r0}[\rho_0]}{d\rho_0}  \xi. \label{delta p_r2} \eeq
If one now identifies the adiabatic "index" as
\beq
\gamma=\frac{\rho_0+p_{r0}}{p_{r0}}\frac{dp_{r0}[\rho_0]}{d\rho_0},
\label{ad index} \eeq
the result derived in Eq.~(\ref{delta p_r2}) reduces to that of
Dev and Gleiser~\cite{Gleiser:Stability}, Eq. (86). However, the
expressions for $\gamma$ differ.
Moreover, if one turns off anisotropy ($\Pi_0=0$) Chandrasekhar's
result is obtained.

\section{ The pulsation equation as an eigenvalue problem}
As in the Chandrasekhar method all matter and metric functions
exhibit oscillatory behavior in time, $f(r,t)=e^{i\omega t}f(r)$.
Hence the pulsation equation assumes the form:
\beq
\mathcal{P}_0\xi ^{\prime \prime}+\mathcal{P}_1\xi^\prime+\mathcal{P}_2\xi=-\omega^2\mathcal{P}_{\omega} \xi,
\label{diff eq}
\eeq
where $\mathcal{P}_0, \mathcal{P}_1, \mathcal{P}_2$ and
$\mathcal{P}_\omega$ are polynomial functions of $r$, depending on
the static solutions only (see Fig. 3). Eq.~(\ref{diff eq})
represents an eigenvalue equation for the radial displacement
$\xi$ (with $\omega^2$ being an eigenvalue). Solutions of this
differential equation are obtained by specifying boundary
conditions in the center and at the surface of the gravastar:
\beq \xi=0\;\;\;\mbox{at}\;\;\; r=0,
\label{BC 1}
\eeq
\beq
\Delta p_r=0 \;\;\;\mbox{at}\;\;\; r=R.
\label{BC 2a}
\eeq
The boundary condition in the center demands that there is no displacement
of the fluid in the center of the gravastar.
 The boundary condition at the surface follows from the
requirement that the Lagrangian radial pressure perturbation has
to vanish at the surface~\cite{MTW,Catalogue,Kokkotas:Oscillation} .
In the model presented here where $\Delta
p_r=(dp_{r0}[\rho_0]/d\rho_0) \Delta\rho$, the  sound velocity vanishes at
the surface, $dp_{r0}/d\rho_0|_{r=R}=0$. This means that, apart
from being finite, there are no further restrictions on
$\Delta\rho(R)$. This also implies that it is sufficient to demand
that $\xi(R)$ and $\xi^\prime(R)$ are bounded in order to satisfy
the boundary condition at the surface \cite{Catalogue}. The choice
\beq
\xi^\prime(R)=0,
\label{BC 2b}
\eeq
 enables one to compare the results in the gravastar's atmosphere with the radial oscillations of the
 polytropes. This can be relevant as the EoS of the gravastar's atmosphere
 close to the surface can be approximated by the polytropic
 EoS $p_r \propto \rho^{1+1/n_p}$, where $n_p$ is a polytropic index~\cite{DeBenedictis,Horvat:Charged}.\\
 %
 In order to study radial stability of the system described by Eq.~(\ref{pulsation eq}) subject to the boundary
 conditions~(\ref{BC 1}) and~(\ref{BC 2a}), it is plausible
 to recast the pulsation equation into the standard
Sturm-Liouville form (see \emph{e.g.} Ref.~\cite{MTW}):
 \beq
 (\mathcal{P}\xi^\prime)^\prime+\mathcal{Q}\xi=-\omega^2\mathcal{W}\xi,
 \label{standard SL}
 \eeq
 where
 \beq
 \mathcal{P}=e^{\int\mathcal{P}_1/\mathcal{P}_0\;  dr}\qquad\mbox{and}\qquad \mathcal{Q}=\frac{P_2}{P_0}P,\;\;\;\; \mathcal{W}=\frac{P_\omega}{P_0}\mathcal{P}.
 \eeq
The leading coefficient in the pulsation equation $\mathcal{P}_0$ has three zeros - two at the ends $\{0,\,R\}$
 and one in the interior region $r_0$ ($dp_{r0}/d\rho_0|_{r=r_0}=0$), hence $\mathcal{P}_1/\mathcal{P}_0$
 has three singular points (see Fig. 3), though all three are regular singular points or Fuchsian
 singularities~\cite{Slavyanov}~\footnote{A singular point $r^*$ is regular (or Fuchsian) if the function $\mathcal{P}_1/\mathcal{P}_0$
 has a pole of at most first order, and the function $\mathcal{P}_2/\mathcal{P}_0$ has a pole of at most second order
 at the singular point $r=r^*$.}.\\

\begin{figure}
\begin{center}
\leavevmode
\includegraphics[scale=0.9]{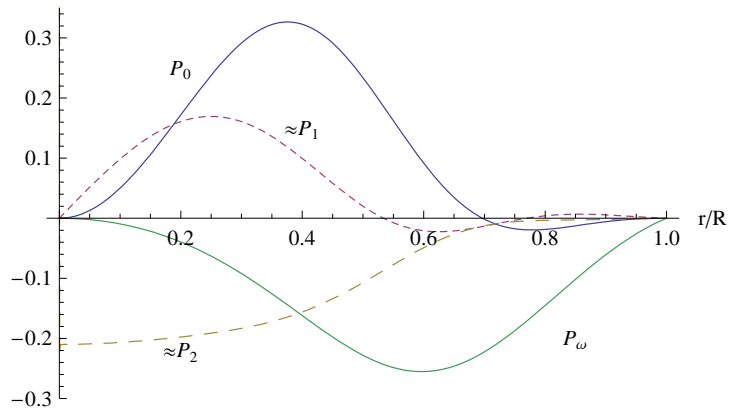}
\end{center}
\caption{The polynomial functions $\mathcal{P}_0,
\mathcal{P}_1/20, \mathcal{P}_2/100$ and $\mathcal{P}_\omega$ from
the pulsation equation~(\ref{pulsation eq}) for
$\{R,n,m\}=\{1,2,3\}$ and $\{\rho_c,\beta\}=\{0.202,81.410\}.$}
\label{fig3}
\end{figure}

In order to obtain $\mathcal{P}$ the integral
$\int \mathcal{P}_1/\mathcal{P}_0\,dr$ should be calculated, and since the interior
singularity arises at $r_0$ which is a division point between
propagating and non-propagating domains, it is reasonable to
divide
 the whole interval $\mathcal{I}=(0,R)$ in two parts:  $\mathcal{I}_1=(0,r_0)$ and
 $\mathcal{I}_2=(r_0,R)$. In performing the integration numerically infinitesimally small regions around all three
 singular points $\{0,r_0,R\}$ are excluded, so that both integrals are rendered convergent and finite. As a consequence, the leading coefficient
 in the Sturm-Liouville equation $\mathcal{P}$ is a positive function on the (whole) interval
 $\mathcal{I}$, whilst the weight function $\mathcal{W}$ is negative on the interval $\mathcal{I}_1$ and
 positive on the interval $\mathcal{I}_2$.
 As elucidated in the previous section, the interesting region is
 the gravastar's atmosphere, \emph{i.e.} the second
 interval, $\mathcal{I}_2$.
 In this region the standard Sturm-Liouville eigenvalue problem formalism (see \emph{e.g.} \cite{Catalogue})
 is applied, since $\mathcal{P}>0$ and $\mathcal{W}>0$. Therefore if $\omega^2$ is
 positive, $\omega$ itself is real and the solution is oscillatory.
  If on the other hand  $\omega^2$ is negative, $\omega$ is
  imaginary and the solution is exponentially
  growing or decaying in time, thus signalizing
  instabilities. The number of nodes of the eigenvector
  $\xi$ for a given eigenvalue $\omega^2$ is closely
  related to the stability criteria. To be more precise,
  if for $\omega^2=0$ eigenvector $\xi$ has no nodes,
  then all higher frequency radial modes are stable.
  Otherwise, if for $\omega^2=0$ eigenvector $\xi$ exhibits
  nodes, then all radial modes are unstable.
  Furthermore, if the system is stable, then the following relations hold
  \beq
  \omega^2_0<\omega^2_1<\dots<\omega^2_n<\dots,
  \eeq
where $n$ equals the number of nodes.\\

\section{Results and Discussion}
In testing stability of certain configurations in general, it
seems natural that one attempts to find critical values of the
parameters for which the system is marginally stable. Marginal
stability means
 here that there exists a set of parameters for which the system
 exhibits the \emph{stable} fundamental mode ($n=0$) for  $\omega^2_0=0$.
 Then for the given set of parameters all higher frequency
modes are radially stable. For example, in the case of neutron
stars (described by the polytropic EoS), there exists a critical
value of the central energy density for which the stellar mass $M$
as a function of radius $R$ is extremal. For such a critical value
of the central energy density the star exhibits stable fundamental
mode with $\omega_0^2=0$, which implies that all higher frequency
modes with the given central energy density are then radially
stable. Furthermore, at the account of the $M(R)$ curve one can
then read off which configuration of the EoS
will produce a stable star and which will not.\\

\begin{figure}
\begin{center}
\leavevmode
\includegraphics[scale=1]{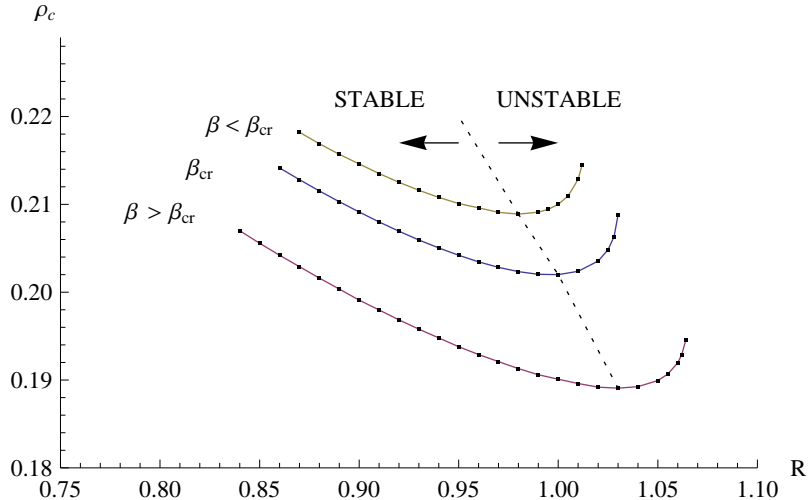}
\end{center}
\caption{The central energy density $\rho_c$ plotted against the
radius $R$. For $\{R,n,m\}=\{1,2,3\}$ the anisotropy-parameter
$\beta=\{92.905,81.410,76.110\}$ is constant on each curve and
fixed by choosing the central energy density to be
$\rho_c=\{0.190,0.202,0.210\}$ from the lower to the upper curve,
respectively. The minimum of each curve represents marginally
stable configurations.}
 \label{Fig4}
\end{figure}

The continuous pressure gravastar model described here displays a
quite similar behavior. For each EoS (fixed $\beta$) the extremum
of the $\rho_c(R)$ curve represents critical values of the
parameters $\{\rho_c,R\}$ for which the system exhibits a stable
fundamental mode, $\omega_0^2=0$ (see Fig. 4). Then for
 such critical set of parameters all higher frequency modes are radially stable.
 Moreover, for smaller radii the system exhibits stability, whereas for
 larger radii (than the critical one) it reveals instability (see Fig. 4).
 In this sense the $\rho_c(R)$ curve for the gravastar mimics the
 well known $M(R)$ curve for a polytrope.
To prove these statements, the behavior of the displacement
function $\xi$ is shown in Fig. 5 for three different cases. The
radius $R$ is, for simplicity, fixed (for all three cases) to be
the critical radius of the central curve in Fig. 4. According to
Fig. 4 one then expects that the radial displacements $\xi$
derived from the lower, central and upper curve in Fig. 4 will
generate stable, marginally stable and unstable fundamental mode
respectively. This is exactly what is shown in Fig. 5.
 The central (solid) curve in Fig. 5 represents
the marginally stable fundamental mode -- an eigenvector $\xi$ is
obtained for $\omega_0^2=0$. The upper (short-dashed) curve
  clearly shows stability of all radial modes as for $\omega^2_0=0$
  there are no nodes,  while the lower (long-dashed) curve reveals
  instabilities of all radial modes as there is a node
  in the fundamental mode. The lower, middle and upper curves in Fig. 5
  correspond to the upper, middle and lower  curves in Fig. 1 and
  Fig. 2, respectively. Here again one is able to relate this result
  to that of Ref.~\cite{Gleiser:Stability}: from Fig. 5,
  according to the values of the anisotropy strengths $\beta$,
  one can conclude that the anisotropy enhances  stability.\\
  Albeit from the viewpoint of radial pulsations, the gravastar's inner region
  does not seem to be physically attractive as the sound velocity is imaginary there, it is important to
  add a couple of comments on the radial displacement's behavior in that region. Pulsations are
  strongly attenuated in the gravastar's interior (see Fig. 5). This holds
  for all $\omega^2>0$. Therefore the radial pulsations of the gravastar as a whole
   can be seen as occurring prevalently in the gravastar's atmosphere whereas entering
   the interior region they are exponentially (but smoothly) attenuated. This is actually what
   one would intuitively expect from the repulsive gravitation caused by the de Sitter-like
   interior.~\footnote{A good example of such a space is an inflationary universe.
   The electric and magnetic fields of free photons in such an inflationary (quasi-de Sitter)
   space get (exponentially) damped as $\propto 1/a^2$, while the physical wavelength
   gets stretched as $\propto a$. Here $a$ denotes the scale factor of the Universe,
   which during inflation grows nearly exponentially in time.}

\begin{figure}
\begin{center}
\leavevmode
\includegraphics[scale=1]{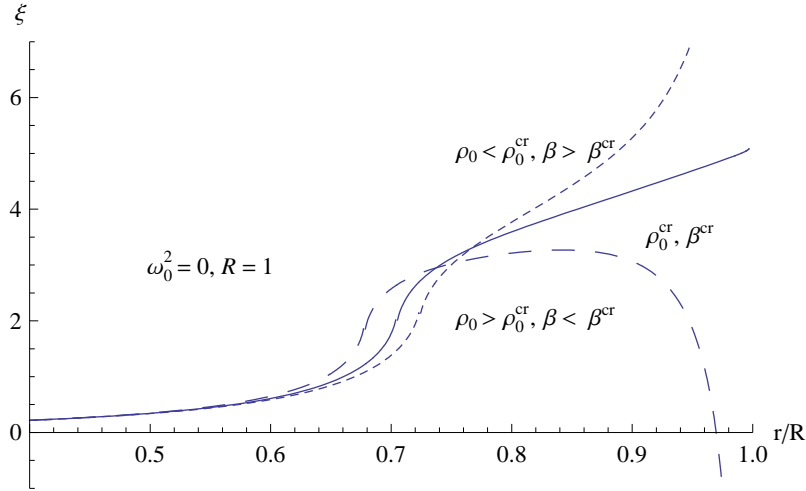}
\end{center}
\caption{The displacement function $\xi(r)$ for
$\{R,n,m\}=\{1,2,3\}$ and $\omega^2=0$. Three different values of
the central energy density $\rho_c=\{0.190,0.202,0.210\}$ and
their respective anisotropy strengths
$\beta=\{92.905,81.410,76.110\}$ correspond to the lower
(unstable), middle (marginally stable) and upper (stable) curve
respectively.}
 \label{fig5}
\end{figure}

In this paper the focus was set on one very specific star model - the
gravastar. Therefore
standard stability analysis which has been applied here  in every
detail could be considered
as a toy model of the radial stability analysis.  However it could be
extended to a broader class of
anisotropic stars with the
anisotropy being a functional of the energy density.
In this way the
adiabatic "index" does not have to be set to a constant but
calculated from the static configurations. This comprises one of
the main results of this paper.

The main result of this work is the observation that the
continuous pressure gravastar-model presented here exhibits radial
stability as illustrated in Fig. 4 and Fig. 5. This result is
important as it, along with the axial stability analysis, suggests
that gravastars, although not yet fully understood at the
fundamental level, may be viable physical compact object
candidates.

\section*{Acknowledgements}
The authors would like to thank Andrew DeBenedictis and Tomislav
Prokopec for useful comments on the manuscript. This work is
partially supported by the Croatian Ministry of Science under the
project No. 036-0982930-3144.

\end{document}